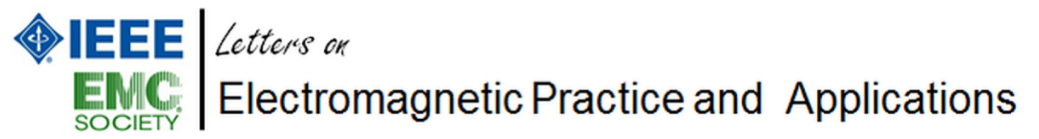

# EFFECT OF ADDING WAVE DIFFRACTORS WITHIN REVERBERATION CHAMBERS ON THE FREQUENCY SPACING OF ADJACENT RESONANT MODES

François Sarrazin, and Guillaume Andrieu, *Senior Member, IEEE*

***Abstract*—This paper takes advantage of a recent method able to extract the characteristics of resonant modes in a metallic enclosure such as a reverberation chamber (RC). The aim here is to analyze, the effect of inserting curvilinear objects within a parallelepiped RC on the chamber performances, particularly from the point of view of the frequency spacing of adjacent resonant modes. Two configurations are compared: one is a parallelepiped RC with added curvilinear diffracting objects, and the other is the same chamber without diffractors but with added absorbers to compensate the decrease of the quality factor. The obtained results exhibit differences that fall within the measurement uncertainties.**

***Index Terms*—chaotic cavity, matrix Pencil, pole extraction, reverberation chamber, scattering parameters.**

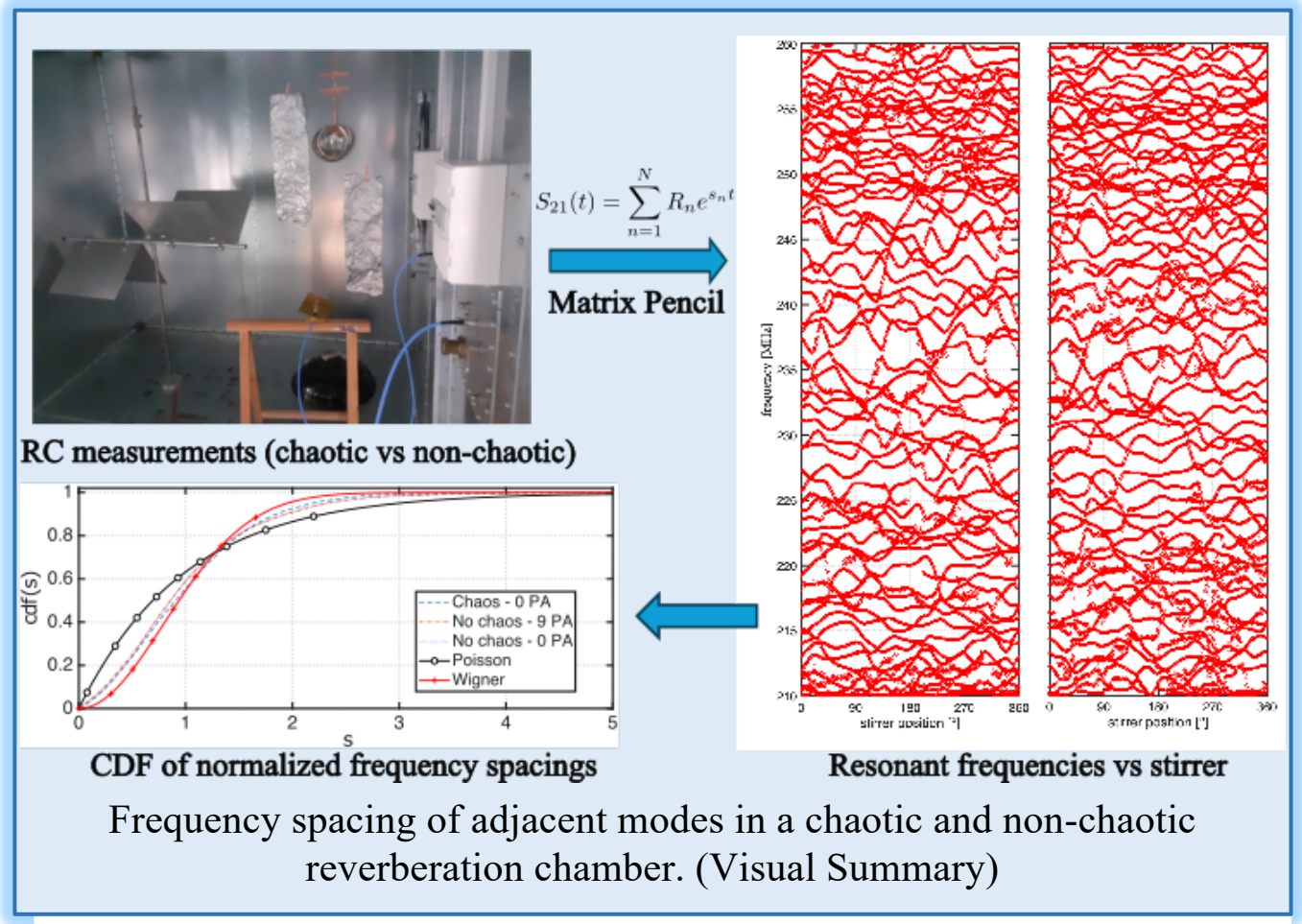

Frequency spacing of adjacent modes in a chaotic and non-chaotic reverberation chamber. (Visual Summary)

## I. INTRODUCTION

Reverberation chambers (RCs) are now widely used for various applications, including electromagnetic compatibility testing and antenna characterization [1]. These tests rely on the assumption of spatial field uniformity within the working volume, which is achieved through various stirring processes. Inspired by wave chaos theory, the concept of a chaotic cavity was first introduced within the RC community in the late 1990s [2]. In particular, the goal is to enhance the RC statistical properties at low frequencies by adding curved wave diffractors, as shown analytically in [3]. Since then, while some contributions have clearly demonstrated such an improvement via numerical simulations [4,5], the practical enhancement of these statistical properties brought about by inserting diffracting metallic objects, such as hemispheres, remains a controversial topic [6]. For example, no clear differences were found in the measured statistical properties of chaotic (with curved wave diffractors) and non-chaotic (without curved wave diffractors) RCs in [7,8]. Still, results may depend on the criterion considered to compare RC configurations. In this letter, we focus on the statistics of the frequency spacing of adjacent resonant modes, also referred to as nearest-neighbor spacings [9]. Indeed, this criterion is particularly discriminating, as the distributions are expected to follow a Poisson distribution for parallelepiped cavities (also called ``integrable" as the field can be computed analytically) whereas a Wigner distribution is expected for chaotic ones. This behavior has been validated using numerical simulations in [10], through the comparison between an integrable cavity and the same cavity in which a corner has been replaced with a sp herical shape. Another study [11] focused on the comparison between a regular RC equipped with a stirrer and the same RC in which the stirrer has been replaced by three hemispheres, leading to statistical distributions closer to Wigner in the case using hemispheres. However, measurement results were not analyzed.

The objective of this letter is to experimentally evaluate the benefits of adding wave diffractors through the analysis of the nearest-neighbor spacing metric. Therefore, it provides an experimental comparison of this metric measured in a regular parallelepiped RC (non-chaotic) and the same RC with added metallic diffractors (chaotic). As it is known that the RC $Q$-factor is

***Take-Home Messages:***

- Chaotic reverberation chambers are a controversial topic in which the practical improvement brought about by adding wave diffractors has yet to be demonstrated.
- The normalized frequency spacing between adjacent resonant modes is a relevant metric for assessing the chaotic behavior of an RC.
- The addition of wave diffractors in a regular parallelepiped RC produces only a weak effect on this metric in our measurement results.
- The Weyl formula has been experimentally validated in a 19 m$^3$ RC in the 210-260 MHz frequency range (approximately 60 modes).

This work is supported in part by the French ``Agence Nationale de la Recherche" (ANR) under Grant ANR-22-CPJ1-0070-01.
François Sarrazin is with Univ Rennes, INSA Rennes, CNRS, IETR-UMR 6164, F-35000 Rennes. (e-mail: francois.sarrazin@univ-rennes.fr).
Guillaume Andrieu is with the XLIM laboratory, 123 avenue Albert Thomas, 87000 Limoges, France (e-mail: guillaume.andrieu@xlim.fr).

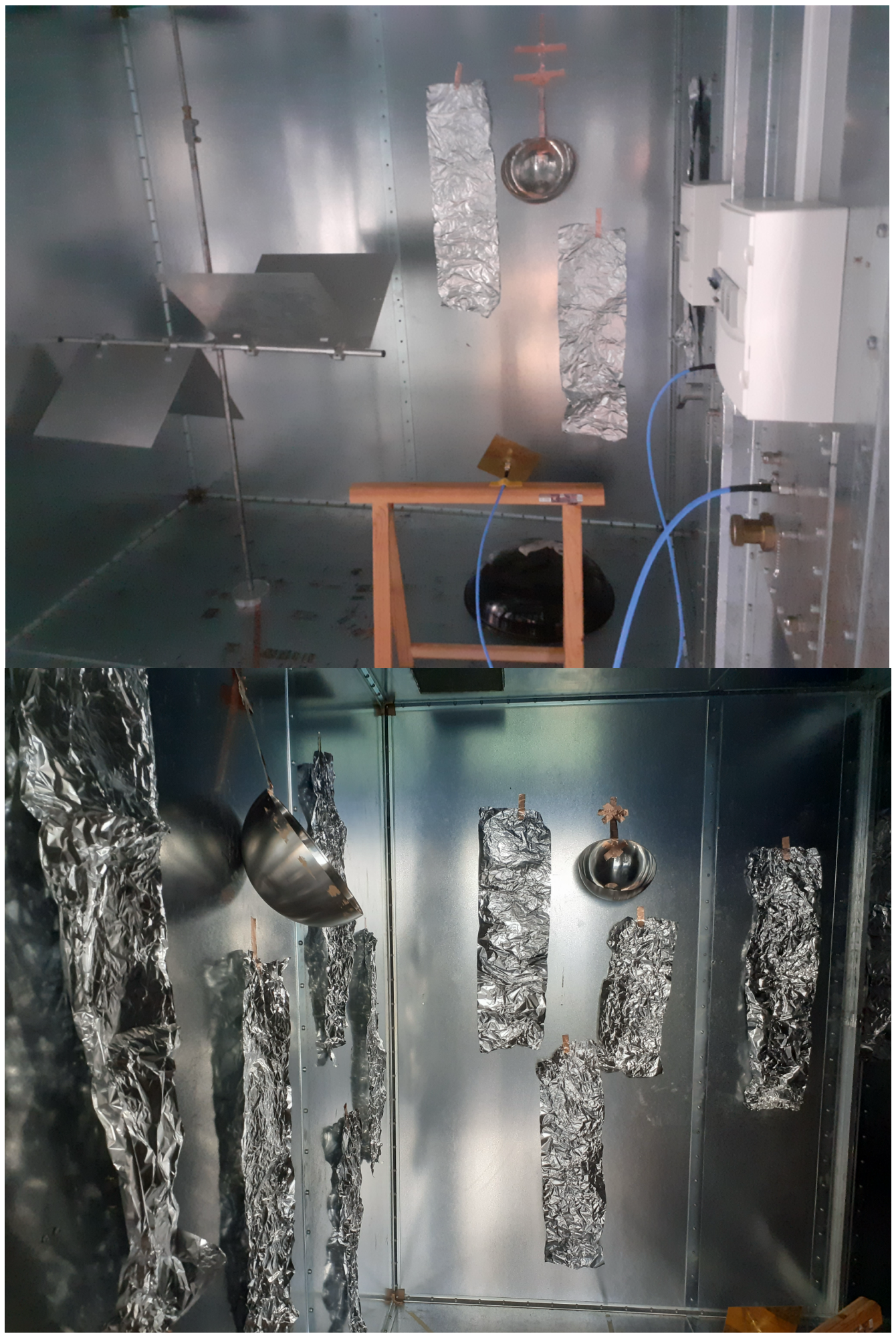

Fig. 1. Two pictures of the chaotic RC. On the top one, we can see the 4-plate mode stirrer, the ground plane of a monopole antenna, a large black circular metallic cover, and 1 hemispherical shell. On the bottom figure, 9~corrugated aluminum strips are visible as well as two hemispherical shells.

degraded by the presence of metallic diffractors [7] due to the increased metallic surface area, this decrease in $Q$-factor must be compensated for by inserting electromagnetic absorbers into the regular RC to enable an unbiased comparison. Resonant modes are extracted from scattering parameter measurements using the Matrix Pencil (MP) algorithm [12,13]. This method has already been successfully applied to mode extraction in RCs [14]. The presented modal analysis also provides an experimental validation of the Weyl's formula [15] regarding the number of modes resonating in a cavity of a given volume over a specified frequency range.

This letter is organized as follows. Section II describes the measurement setups for both the non-chaotic and chaotic RCs. After showing that the $Q$-factor is similar for both RC configurations, thanks to the addition of absorbers in the non-chaotic case, Section III summarizes the basic principles of the extraction method and presents the results in terms of mode tracking and nearest-neighbor spacing distributions.

## II. Description of the Measurements

Measurements related to the results shown in this paper were performed in the RC of the XLIM laboratory (3.75 m long, 2.45 m wide, 2.46 m high, volume V ≈ 22.6 m$^3$, theoretical fundamental resonance $f_0$ around 73 MHz) shown in Fig. 1. Several aspects of the experimental setup have been chosen to 1) increase the likelihood of observing differences between the two RC configurations, and 2) reliably track the modes as the stirrer rotates. First, the RC is equipped with a rotating mode stirrer made of only 4 rectangular metallic plates (60 by 40 cm$^2$) [16]. Four plates were removed from the original stirrer to minimize perturbations of the cavity modes and improve tracking using the MP method. In addition, 692 mode stirrer positions were considered, corresponding to a rotation of 0.52∘ between successive positions. Second, the frequency range was selected to ensure a low modal density, i.e. 210-260 MHz over 10001 linearly spaced frequencies ($\delta f = 5$ kHz), allowing a manageable number of modes. In this range (below the theoretical lowest usable frequency of the RC), the chamber is known to be poorly stirred [17], regardless of the Q-factor. Third, two monopole antennas made of a 40-cm metallic wire mounted orthogonally on a 15 cm square ground plane were inserted into the RC and connected to a vector network analyzer (VNA) via coaxial cables. Their dimensions were chosen to ensure good matching at the central frequency while minimally disturbing the field distribution.

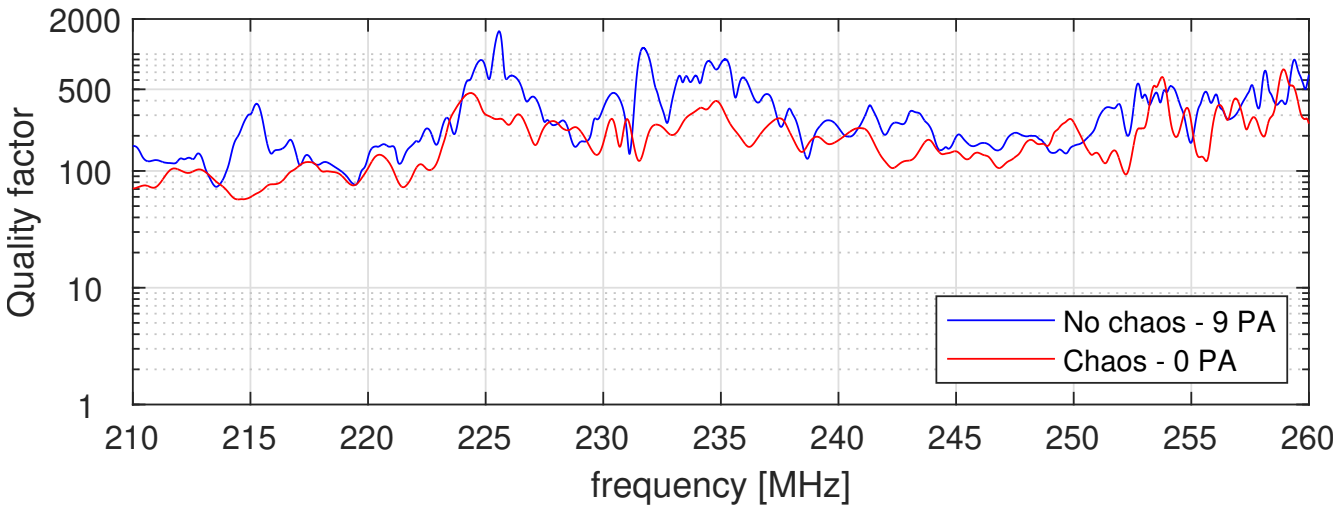

Fig. 2. Average composite $Q$-factor calculated using [19] in the 210~MHz to 260~MHz bandwidth for the non-chaotic and the chaotic configurations

The objective is to compare chaotic and non-chaotic configurations. The non-chaotic configuration corresponds to the parallelepiped RC including the mode stirrer and both antennas. For the chaotic configuration, several objects were inserted into the RC: a metallic sphere (28 cm diameter, 0.22 λ at the central frequency), three hemispherical shells of the same diameter (all made of stainless steel), 15 corrugated aluminum strips (1 m × 27 cm, i.e. $0.78\,\lambda \times 0.21\,\lambda$), and a circular metallic cap of 59 cm diameter (λ/2) with a maximum height of 18 cm (0.14 λ). These objects were installed on the walls and floor to prevent parallel walls from directly facing each other [18]. To ensure a fair comparison, both configurations must have the same average composite Q-factor. However, inserting metallic diffractors slightly alters the RC Q-factor [7]. Therefore, in the non-chaotic configuration, a block of nine pyramidal absorbers (PA) was placed below the mode stirrer to compensate for the additional metallic losses. Each PA has a square base of 100 cm$^2$ and a height of 30 cm. Under these conditions, the measurement time for each configuration is about 160 minutes (after VNA calibration). The RC Q-factor is computed using (6) in [19] from the measured S-parameters, with results shown in Fig. 2. The Q-factor is similar for both configurations, enabling a fair comparison.

## III. Results

### A. Mode Extraction

The method presented extensively in [14] is here briefly summarized. It is based on transfer function measurements (i.e. $S_{21}$) between two antennas inserted in the RC for different stirring conditions (i.e. different mode stirrer positions for our mechanically stirred RC). An inverse Fourier transform is then applied on each frequency-domain measurement made for a given stirring condition. Finally, the time-domain transmission coefficient $S_{21}(t)$ is modelled as an exponential sum as follows

$$S_{21}(t) = \sum_{n=1}^{N} R_n e^{s_n t} \quad (1)$$

where $s_n = \sigma_n + j2\pi f_n$ is the $n^{\text{th}}$ pole, $\sigma_n$ the damping coefficient, $f_n$ the resonant frequency, $R_n$ the complex residue associated to the $n^{\text{th}}$ pole, and $N$ the total number of poles. According to Weyl's formula, about 58 modes are expected in the considered frequency range. Therefore, $N$ was set to 150 in this work to slightly overestimate the expected number of modes (two complex-conjugate poles are extracted per mode). This overestimation is required to retrieve all the true (physical) poles as spurious poles are also extracted due to the presence of noise [14]. The frequency $f_n$ of each pole is presented as a function of the stirrer position in Fig. 3 for both RC configurations, where each frequency is represented by a red circle. Due to the small between adjacent stirrer positions (0.52° rotation step), these markers visually form continuous lines, allowing each extracted mode to be tracked along the stirring process. While providing insight into the absence of a clear difference (at first sight) in mode variation with stirrer rotation for both RCs, this figure also validates the quality of the mode extraction through the clear continuity of the modes. However, in addition to the continuous lines corresponding to true modes, a large number of isolated spurious poles is randomly extracted by the MP algorithm. Although dedicated post-processing techniques based on time stability have been developed to filter out these spurious poles, the fine stirrer step in this specific experiment allows filtering based on pole continuity with respect to stirrer position. The filtering approach was therefore carried out in two steps: 1) a systematic procedure in which, for each pole of a stirrer position, the presence of a nearby pole at the previous and next positions is checked; otherwise, the pole is considered isolated and removed; 2) the remaining poles not eliminated in the first step but clearly spurious are removed manually, i.e. one by one. It is noticed that the same approach was performed for both measurement sets. The frequencies of all poles retained after filtering are presented in Fig. 4. Only true modes remain, as the pole continuity is clearly visible. The number of excited modes is presented as a function of the mode stirrer position in Fig. 5. The number of modes is very similar for both configurations (average difference of 1.56 modes) and varies only slightly with stirrer position (standard deviations of 1.89 and 2.98 for the non-chaotic and chaotic cases, respectively). Finally, the number of modes retrieved experimentally is close to that predicted by Weyl's formula (average differences of 2.88 for the non-chaotic case and 4.45 for the chaotic case). To the best of the authors' knowledge, these results represent the first experimental validation of Weyl's formula in the RF domain.

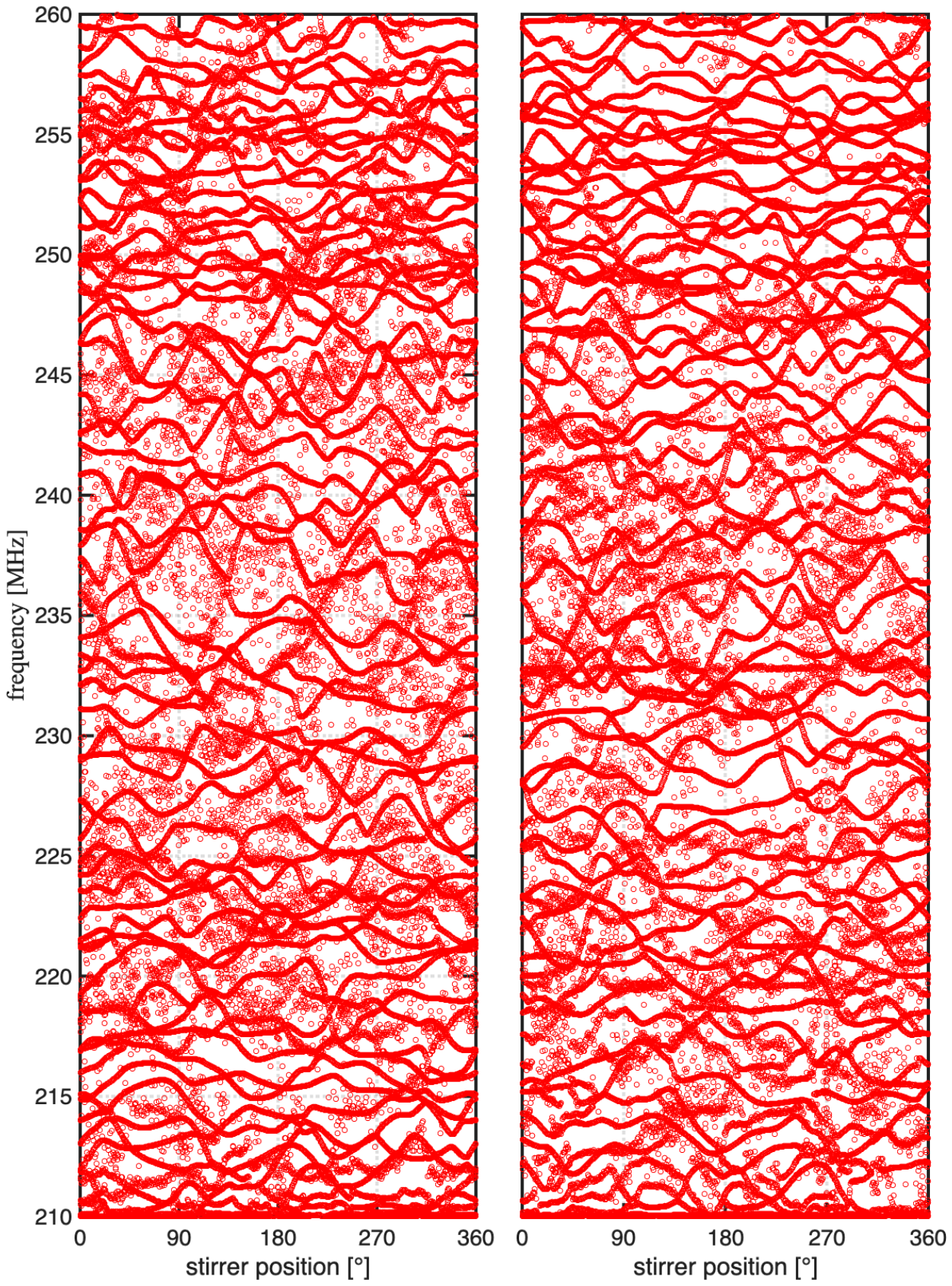

Fig. 3. Frequency $f_n$ of each pole as a function of the stirrer position before filtering for the non-chaotic (left) and chaotic (right) configurations.

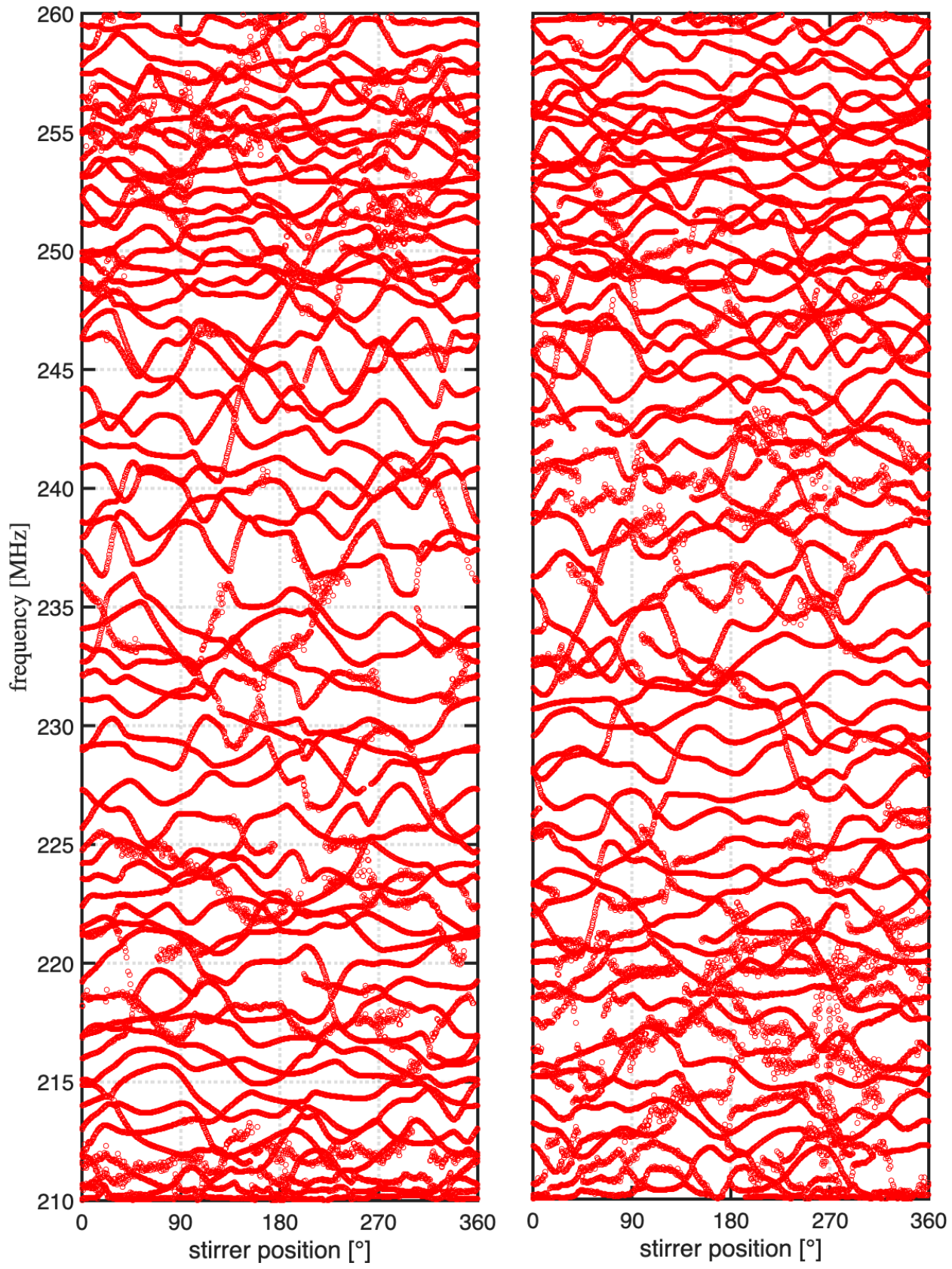

Fig. 4. Frequency $f_n$ of each pole as a function of the stirrer position after filtering for the non-chaotic (left) and chaotic (right) configurations.

## B. *Analysis of the Frequency Spacing Distribution*

The cumulative distribution function (CDF) of the normalized frequency spacing between adjacent modes s [4] (data from all the mode stirrer positions are considered here) is presented in Fig. 6.

They are compared with the Wigner CDF (which should be obtained in a chaotic RC) and the Poisson CDF (which should be obtained in a parallelepiped RC). As mentioned before on other plots, the difference between both RC configurations is difficult to observe. To quantify the difference, we fit the frequency spacing distribution to the Brody's formula [20] given by P Brody $q(s) = asq\,exp(-bsq+1)$ where $a = (q+1)b$, $b = \Gamma\{(q+2)/(q+1)\}q+1$, $\Gamma(n) = (n-1)!$ and $q$ is the level repulsion parameter. If $q = 0$, then (2) reduces to a Poisson distribution whereas if $q = 1$, then (2) becomes a Wigner distribution. Results are presented in Table I. The uncertainty has been computed as 5 % above the minimum q value obtained from the fitting process. While a slight improvement is observed for the chaos case ($q$ closer to 1), the difference remains within the measurement uncertainties. In addition, we performed the same study for the non-chaotic case without adding absorbers (No Chaos - 0 PA). It leads to the exact same value of $q$, confirming that this indicator does not depend on the $Q$-factor. Also, it is interesting to notice that all three curves look like more to a Wigner distribution than to a Poisson one. This suggests that the presence of a mechanical stirrer, antennas, and small imperfections (for instance screws) already has an impact on the RC statistical properties and that adding diffracting objects inside the RC does not introduce a noticeable difference.

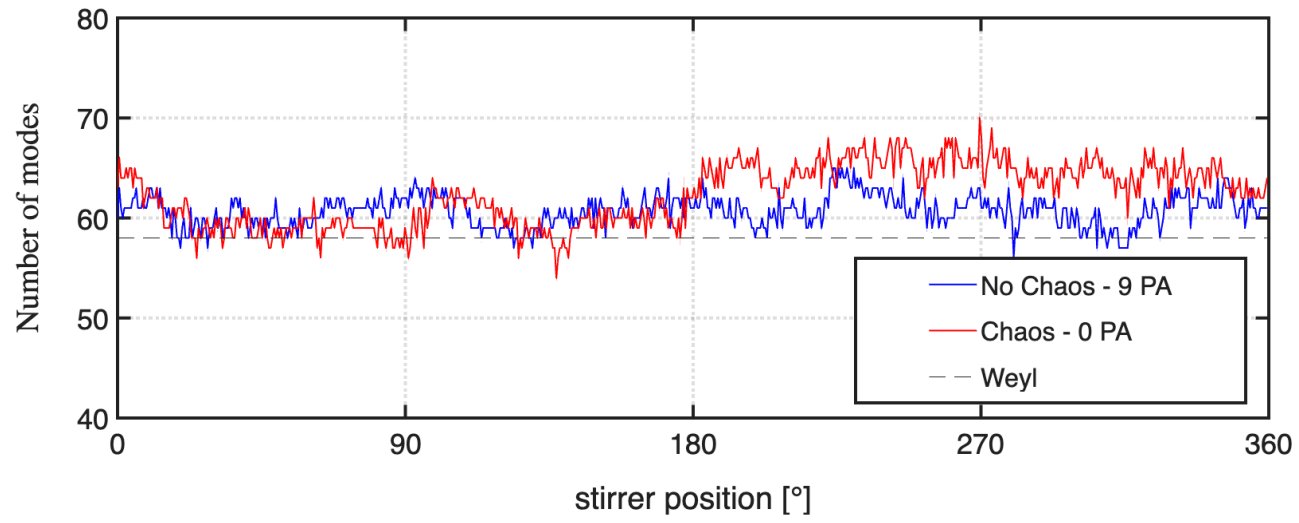

Fig. 5. Number of modes extracted for each configuration as a function of the mode stirrer position after filtering. Comparison with the Weyl's formula.

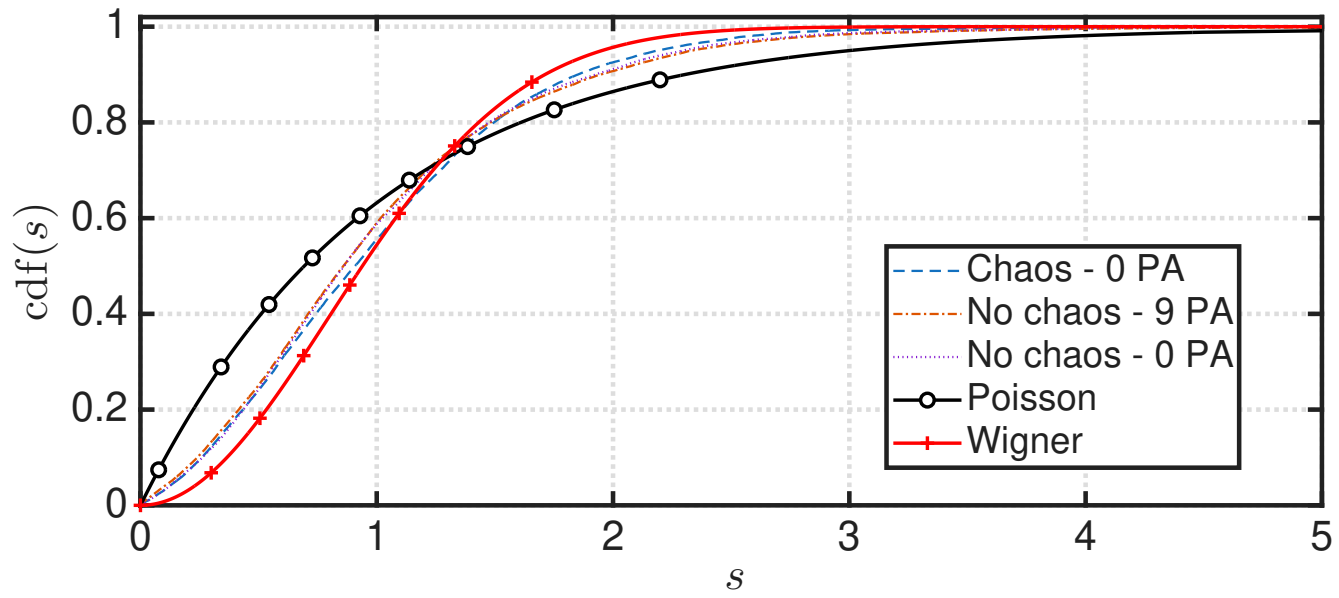

Fig. 6. CDF of the normalized frequency spacing between adjacent frequencies. Comparison with the Wigner and the Poisson distribution.

Table 1: Brody's $q$ parameter for each RC configuration

| RC configuration | Brody's $q$ parameter |
|---|---|
| *No chaos – 9 PA* | $0.56 \pm 0.03$ |
| *Chaos – 0 PA* | $0.61 \pm 0.04$ |
| *No chaos – 0 PA* | $0.56 \pm 0.04$ |

## IV. Conclusion

This letter builds on a previously published method for extracting the resonant modes of an RC from scattering parameter measurements to investigate the effect of adding curvilinear diffracting objects on RC performance. The comparison was carried out under identical Q-factor conditions, with absorbers inserted to compensate for the Q-factor reduction caused by the diffracting objects. Following an experimental validation of Weyl's formula, the results show that the effect of these objects on the statistics of the frequency spacing between adjacent modes is weak and remains within measurement uncertainties. These conclusions, drawn for the first time from the experimental analysis of the nearest-neighbor spacing metric, are consistent with previous works [7,8].

## REFERENCES

[1] G. Andrieu, Electromagnetic Reverberation Chambers - Recent advances and innovative applications. IET, 2020.

[2] L. Cappetta, M. Feo, V. Fiumara, V. Pierro, and M. Pinto, "Electromagnetic chaos in mode-stirred reverberation enclosures," IEEE Trans. Electromagn. Compat., vol. 40, no. 3, pp. 185–192, Aug. 1998.

[3] L. Arnaut, "Operation of electromagnetic reverberation chambers with wave diffractors at relatively low frequencies," IEEE Trans. Electromagn. Compat., vol. 43, no. 4, pp. 637–653, Nov. 2001.

[4] J.-B. Gros, O. Legrand, F. Mortessagne, E. Richalot, and K. Selemani, "Universal behaviour of a wave chaos based electromagnetic reverberation chamber," Wave Motion, vol. 51, no. 4, pp. 664–672, 2014.

[5] J.-B. Gros, U. Kuhl, O. Legrand, F. Mortessagne, O. Picon, and E. Richalot, "Statistics of the electromagnetic response of a chaotic reverberation chamber," Adv. Electromagn., vol. 4, no. 2, p. 38, 2015.

[6] R. Serra et al., "Reverberation chambers at the edge of chaos: Discussion forum at emc europe 2020," IEEE Electromagn. Compat. Mag., vol. 11, no. 1, pp. 73–88, 2022.

[7] G. Andrieu, "On the possible benefits of inserting metallic diffractors to improve low frequency performance of reverberation chambers," IEEE Trans. Electromagn. Compat., vol. 63, no. 1, pp. 304–307, 2021.

[8] M. Magdowski, J. Immidisetti, and R. Vick, "Experimental analysis of the field homogeneity and isotropy inside a reverberation chamber with two hemispherical diffractors," in Int. Symp. Electromagn. Compat. (EMC EUROPE), Aug 2018, pp. 683–688.

[9] A. Cozza, "Probability distributions of local modal-density fluctuations in an electromagnetic cavity," IEEE Trans. Electromagn. Compat., vol. 54, no. 5, pp. 954–967, 2012.

[10] G. Orjubin, E. Richalot, O. Picon, and O. Legrand, "Chaoticity of a reverberation chamber assessed from the analysis of modal distributions obtained by fem," IEEE Trans. Electromagn. Compat., vol. 49, no. 4, pp. 762–771, 2007.

[11] E. Richalot et al., "Criterion based on resonant frequencies distributions for reverberation chamber characterization," in 2015 Int. Conf. Electromagn. Advanced Appl. (ICEAA), 2015, pp. 1112–1115.

[12] Y. Hua and T. Sarkar, "Matrix pencil method for estimating parameters of exponentially damped/undamped sinusoids in noise," IEEE Trans. Acoust., Speech, Signal Process., vol. 38, no. 5, pp. 814–824, 1990.

[13] F. Sarrazin, J. Chauveau, P. Pouliguen, P. Potier, and A. Sharaiha, "Accuracy of singularity expansion method in time and frequency domains to characterize antennas in presence of noise," IEEE Trans. Antennas Propag., vol. 62, no. 3, pp. 1261–1269, 2014.

[14] F. Sarrazin and E. Richalot, "Accurate characterization of reverberation chamber resonant modes from scattering parameters measurement," IEEE Trans. Electromagn. Compat., vol. 62, no. 2, pp. 303–314, 2020.

[15] H. Weyl, " ¨Uber die randwertaufgabe der randwertaufgabe der strahlungstheorie und asymptotische spektralgesetze," J. Reine U. Angew. Math., vol. 143, pp. 177–202, 1913.

[16] A. Adardour, G. Andrieu, and A. Reineix, "On the low-frequency optimization of reverberation chambers," IEEE Trans. Electromagn. Compat., vol. 56, no. 2, pp. 266–275, April 2014.

[17] G. Andrieu, N. Ticaud, F. Lescoat, and L. Trougnou, "Fast and accurate assessment of the "well stirred condition" of a reverberation chamber from S11 measurements," IEEE Trans. Electromagn. Compat., vol. 61, no. 4, pp. 974–982, Aug 2019.

[18] K. Selemani et al., "Comparison of reverberation chamber shapes inspired from chaotic cavities," IEEE Trans. Electromagn. Compat., vol. 57, no. 1, pp. 3–11, 2015.

[19] P. Besnier, C. Lemoine, and J. Sol, "Various estimations of composite q-factor with antennas in a reverberation chamber," in IEEE Int. Symp. Electromagn. Compat. (EMC), Aug 2015, pp. 1223–1227.

[20] T. A. Brody, J. Flores, J. B. French, P. A. Mello, A. Pandey, and S. S. M. Wong, "Random-matrix physics: spectrum and strength fluctuations," Reviews of Modern Physics, vol. 53, no. 3, pp. 385–479, Jul. 1981.